\documentclass[twocolumn,
,showpacs,amssymb,amsmath,aps,pra]{revtex4-1}
\newcommand{\half}{\mbox{$\frac{1}{2}$}}
\usepackage{bm}
\usepackage{graphics}
\usepackage{amsmath}
\usepackage{amssymb}
\usepackage{pstricks}
\usepackage{multido}
\usepackage{epsfig}

\begin{document}
\title{Evaluation of ground state entanglement in spin systems
       with the random phase approximation}
\author{J.M. Matera, R. Rossignoli, N. Canosa}
\affiliation{Departamento de F\'{\i}sica-IFLP,
 Universidad Nacional de La Plata, C.C. 67, La Plata (1900) Argentina}
\date{\today}

\begin{abstract}
We discuss a general treatment based on the mean field plus random phase
approximation (RPA) for the evaluation of subsystem entropies and negativities
in ground states of spin systems. The approach leads to a tractable general
method, becoming straightforward in translationally invariant arrays. The
method is examined in arrays of arbitrary spin with $XYZ$ couplings of general
range in a uniform transverse field, where the RPA around both the normal and
parity breaking mean field state, together with parity restoration effects, are
discussed in detail. In the case of a uniformly connected $XYZ$ array of
arbitrary size, the method is shown to provide simple analytic expressions for
the entanglement entropy of any global bipartition, as well as for the
negativity between any two subsystems, which become exact for large spin. The
limit case of a spin $s$ pair is also discussed.
\end{abstract}
\pacs{03.67.Mn, 03.65.Ud, 75.10.Jm}
\maketitle
\section{Introduction}
The study of entanglement constitutes one of the most active and challenging 
research areas, being of central interest in the fields of quantum information 
\cite{NC.00} and many-body physics \cite{AF.08}. The concept of entanglement 
has provided a new perspective for analyzing quantum correlations and quantum 
critical phenomena in many particle systems, leading to fundamental results 
and new insights in the field \cite{ON.02,VV.03,AF.08,ECP.10}. Nonetheless, the 
evaluation of entanglement in general strongly interacting many-body systems 
remains a difficult task, particularly in systems with long range 
interactions, high connectivity and large dimensionality, where usual treatments
such as Quantum MonteCarlo \cite{QTM.97}, DMRG \cite{SWS.03} or matrix
product states \cite{VC.06} become more involved or difficult to implement. In
previous works \cite{CMR.07,MRC.08} we have applied a general mean field plus
RPA treatment to the evaluation of pairwise entanglement (i.e., that between
two elementary components) in spin systems at zero and finite temperature. The
approach was able to capture the main features of the entanglement between two
spins in arrays with $XY$ and $XYZ$ couplings of different ranges, including the
prediction of full range pairwise entanglement in the vicinity of the
factorizing field \cite{MRC.08,AA.06,RCM.08}. The accuracy of the approach was
shown to increase with the interaction range or connectivity.

The aim of the present work is to examine the capability of the previous method 
for predicting, in the ground state of spin systems, the entanglement properties 
of {\it general} subsystems. We will focus on the entanglement entropy of 
arbitrary bipartitions of the whole system, as well as on the negativity between 
{\it any} two subsystems, not necessarily complementary, where the rest of the 
spins play the role of an environment and entanglement can no longer be measured 
through the subsystem entropy. Other measures, like the negativity (an entanglement 
monotone computable for general mixed states \cite{VW.02,ZHSL.98}) have to be 
employed. This type of entanglement has recently received special attention 
\cite{WMB.09,WVB.10,NC.06} since its behavior can differ from that of global 
bipartitions. We will show that the present approximation provides a general 
tractable scheme for evaluating these quantities, becoming analytic in 
translationally invariant systems.

In Section II we present the general RPA formalism, describing the RPA spin
state, the associated bosonic estimation of subsystem entropies and
negativities, the implementation in translationally invariant systems and the
application to a general spin $s$ array with $XYZ$ couplings of arbitrary range
in a transverse magnetic field. Symmetry restoration effects in the case of
parity-breaking mean fields are also discussed. As illustration, we derive in
sec.\ \ref{III} results for a spin $s$ pair and for a fully connected finite 
spin $s$ array, where RPA is able to provide simple full analytic expressions 
for subsystem entropies and negativities, which represent the exact large spin 
limit at any fixed size. Appendix \ref{ApA} discusses the equivalence between the 
spin and the bosonic RPA treatments, whereas appendix \ref{ApB} contains details 
of the analytic results of sec.\ \ref{III}. Conclusions are drawn in IV.

\section{Formalism}
\subsection{RPA for spin systems at $T=0$}
We will consider a general finite system of spins ${\bm s}_i=
(s_{ix},s_{iy},s_{iz})$, connected through general quadratic couplings and
immersed in a magnetic field, not necessarily uniform. The corresponding
Hamiltonian is
\begin{equation}
H=\sum_{i,\mu}{B}^{i\mu}{s}_{i\mu}-\half\sum_{i\neq j,\mu,\nu}
{J}^{i\mu j\nu}s_{i\mu}s_{j\nu}\,,\label{H}
\end{equation}
where $\mu=x,y,z$ and $B^{i\mu}$ are the field components at site $i$. Ising,
$XY$, $XYZ$ ($J^{i\mu\,j\nu}=\delta^{\mu\nu}J_\mu^{ij}$) as well as
Dzyaloshinskii-Moriya ($J^{i\mu,j\nu}=-J^{i\nu,j\mu}$) couplings of arbitrary 
range are particular cases of Eq.\ (\ref{H}).

The first step in the RPA \cite{RS.80} is to determine the mean field ground
state, i.e., the separable state
\[|0\rangle\equiv\otimes_{i=1}^n |0_i\rangle= |0_1\ldots
 0_n\rangle\]
with the lowest energy $\langle H\rangle_0=\langle 0|H|0\rangle$, given by 
\begin{equation}
\langle H\rangle_0=\sum_{i,\mu}{B}^{i\mu}\langle
s_{i\mu}\rangle_0-\half\sum_{i\neq j,\mu,\nu} {J}^{i\mu j\nu}\langle
s_{i\mu}\rangle_0\langle s_{j\nu}\rangle_0\label{Hm}
\end{equation}
where $\langle \bm{s}_{i}\rangle_0=\langle 0_i|\bm{s}_{i}|0_i\rangle$. Each
local state $|0_i\rangle$  can be determined self-consistently as the lowest
eigenstate of the local mean field Hamiltonian
\begin{equation}
{h}_i=\sum_{\mu}\frac{\partial\langle {H}\rangle_0} {\partial\langle
{s}_{i\mu}\rangle_0}{s}_{i\mu}=\bm{\lambda}^i\cdot\bm{s}_i\,,\label{h}
\end{equation}
being then the state with maximum spin $s_i$ directed along $-\bm{\lambda}^i$
(a local coherent state). This leads to the self-consistent equations
\begin{eqnarray}
\lambda^{i\mu}&=&B^{i\mu}-\sum_{j\neq i,\nu}{J}^{i\mu\,j\nu} \langle
{s}_{j\nu}\rangle_0\,,\; \langle
\bm{s}_{i}\rangle_0=-s_i\bm{\lambda}^i/\lambda^i\,,\label{mf}
\end{eqnarray}
where $\lambda^i=|\bm{\lambda}^i|$. Eq.\ (\ref{mf}) can be solved iteratively
starting from an initial guess for $|0_i\rangle$ or $\bm{\lambda}_i$, although
other procedures (like the gradient method) can  be employed. Eq.\ (\ref{Hm})
becomes then $\langle H\rangle_0=
\half\sum_i(\bm{\lambda^i}+\bm{B}^i)\cdot\langle \bm{s}_i\rangle_0$.

Since the form (\ref{H}) is valid for any choice of the local axes, it
is now convenient to choose $z_i$ along $\bm{\lambda}^i$, such that $\langle
s_{i\mu}\rangle_0=-s_i\delta_{\mu z}$ and $\lambda^{i\mu}=\lambda^i\delta^{\mu
z}$, with $\lambda^i>0$. The second step in the RPA is the approximate
bosonization
\begin{equation}s_{i+}\rightarrow \sqrt{2s_i}b^\dagger_{i}\,,\;\;
s_{i-}\rightarrow \sqrt{2s_i}b_i\,,\;\;s_{iz}\rightarrow-s_i+b^\dagger_ib_i\,,
 \label{sb} \end{equation}
where $s_{i\pm}=s_{ix}\pm is_{iy}$ and $b_i$, $b^\dagger_i$ are considered
standard boson operators ($[b_i,b^\dagger_j]=\delta_{ij}$,
$[b_i,b_j]=[b^\dagger_i,b^\dagger_j]=0$), with $|0\rangle\rightarrow
|0_b\rangle$ their vacuum. This bosonization is in agreement with that implied
by the path integral formalism of \cite{CMR.07,MRC.08} for $T\rightarrow
0$, and preserves two of the exact spin commutators exactly ($[{
s}_i^z,{s}_j^{\pm}]=\pm\delta_{ij}{ s}_i^\pm$), the remaining one preserved as
vacuum average ($\langle [{s}_i^-,{s}_j^+] \rangle_0=2s_i\delta_{ij}$). 
It coincides with the Holstein-Primakoff
and other exact bosonizations \cite{HP.58,RS.80,KM.91,VDB.07}
up to zeroth order in $s_i^{-1}$. 

The third step is to replace Eq.\ (\ref{sb}) in the original Hamiltonian
(\ref{H}), neglecting all cubic and quartic terms in $b_i$, $b^\dagger_i$. This
leads to the quadratic boson Hamiltonian
\begin{eqnarray}
{H}^b&=&\langle {H}\rangle_0+\sum_i \lambda^i {  b}^\dagger_i{ b}_i
-\sum_{i\neq j}\Delta_+^{ij}{ b}^\dagger_i{ b}_j +\half(\Delta_-^{ij}{
b}^\dagger_i{ b}^\dagger_j+h.c.)\nonumber\\
&=&\langle {H}\rangle_0-\half\sum_{i}\lambda^{i}+
\half {\cal Z}^\dagger {\cal H}{\cal Z}\,,\label{Hb}\\
{\cal Z}&=&\left(\begin{array}{c}{ b}\\{ b}^\dagger\end{array}\right)\,,\;\;
{\cal H}=\left(\begin{array}{cc}\Lambda-\Delta_{+}&-\Delta_{-}
\\-\bar{\Delta}_{-}&\Lambda-\bar{\Delta}_{+}\end{array}\right),\label{HM}\\
\Delta_{\pm}^{ij}&=&\half\sqrt{s_i s_j}\,[J^{ix\,jx}\pm J^{iy\,jy}
-i(J^{iy\,jx}\mp J^{ix\,jy})],\label{DPM}
\end{eqnarray}
where ${\cal Z}^\dagger=(b^\dagger,b)$ and $\Lambda^{ij}=\lambda^i\delta^{ij}$.
The choice of the mean field axes for the bosonization (\ref{sb}) ensures that
no linear terms in ${b}_{i}$, ${b}^{\dagger}_{i}$ appear in $H^b$, reflecting
the stability of the mean field state $|0\rangle$ with respect to one site
excitations.

The last step is the diagonalization of the bosonic quadratic form (\ref{Hb}),
which is always possible when the hermitian matrix ${\cal H}$ in (\ref{HM})
{\it is positive definite}, i.e., when $|0\rangle$ is a stable vacuum
\cite{RS.80}. ${H}^b$ can then be rewritten as
\begin{equation}
{H}^b=\langle {H}\rangle_0+\sum_{\alpha}\omega^\alpha
{ b'}^\dagger_{\alpha}
{b}'_{\alpha}+\half(\omega^{\alpha}-\lambda^{\alpha})
\,,\label{Hb3}
\end{equation}
where $\lambda^\alpha$ stands for $\lambda^{i}$, $\omega^\alpha$ are the
symplectic eigenvalues of ${\cal H}$, i.e., the positive eigenvalues of the
matrix
\begin{equation}
{\cal M}{\cal H}=
\left(\begin{array}{cc}\Lambda-\Delta_{+}&-\Delta_{-}\\
\bar{\Delta}_{-}&-\Lambda+\bar{\Delta}_{+}\end{array}\right)\,,
\;\;{\cal M}=\left(\begin{array}{cc}1&0\\0&-1 \end{array}\right)
\label{MH}
\end{equation}
whose eigenvalues come in pairs of opposite sign (and which is diagonalizable
with real non-zero eigenvalues when ${\cal H}$ is positive definite), and ${
b'}_\alpha$, ${{b}'}^\dagger_\alpha$ are ``collective'' boson operators related
to the local ones by a Bogoliubov transformation ${\cal Z}={\cal W}{\cal Z}'$,
i.e.,
\begin{equation}\left(\begin{array}{c}{ b}\\{ b}^\dagger\end{array}\right)
={\cal W}\left(\begin{array}{c}{ b}'\\{b'}^\dagger\end{array}\right)\,,\;\;
{\cal W}=\left(\begin{array}{cc}U&V\\\bar{V}&\bar{U}\end{array}\right)
 \label{W}\end{equation}
with $(^{U}_{\bar{V}})_\alpha$ and $(^{V}_{\bar{U}})_\alpha$ the eigenvectors
of ${\cal M}{\cal H}$ associated with the eigenvalues $\omega_\alpha$ and
$-\omega_{\alpha}$ respectively (such that ${\cal W}^{-1}{\cal M}{\cal H}{\cal
W}={\cal M}\Omega$, with $\Omega_{\alpha\alpha'}=
|\omega_\alpha|\delta_{\alpha\alpha'}$). In order to preserve the boson
commutation relations, which can be cast as ${\cal Z}{\cal Z}^\dagger-[({\cal
Z}^\dagger)^{\rm tr}{\cal Z}^{\rm tr}]^{\rm tr}={\cal M}$,  ${\cal W}$ should
satisfy
\begin{equation}
{\cal W}{\cal M}{\cal W}^\dagger={\cal M}\label{WM}
 \end{equation}
which implies also ${\cal W}^\dagger{\cal M}{\cal W}={\cal M}$ and hence ${\cal
W}^\dagger {\cal H}{\cal W}=\Omega$. This entails $U^\dagger V-{V}^{\rm
tr}\bar{U}=0$, $U^\dagger U-{V}^{\rm tr}\bar{V}=I$, which are the natural
orthogonality relations fulfilled by the eigenvectors of (\ref{MH}) with
normalization $(^{U}_{\bar{V}})_{\alpha}^\dagger{\cal
M}(^{U}_{\bar{V}})_\alpha=1$.

The RPA matrix (\ref{HM}) is of dimension $2n\times 2n$, with $n$ the number of
spins. RPA involves then an exponential reduction in the dimension (from
$(2s+1)^n$ to $2n$ for $n$ identical spins). Moreover, in  a translationally
invariant system (sec.\ \ref{TI}), it can be further reduced to $n$ $2\times 2$
matrices, becoming then {\it fully analytic}.

\subsection{The RPA ground state\label{IIb}}
The vacuum of the new bosons ${b}'$ (${b}'_\alpha|0'_b\rangle=0$) is
\cite{RS.80}
\begin{equation}
|0'_b\rangle=C_b\exp[\half\sum_{i,j}Z^{ij}{b}^\dagger_{i}{b}^\dagger_{j}]
 |0_b\rangle\,,\;\;Z=V\bar{U}^{-1}\,,\label{stb}
 \end{equation}
where $C_b=\langle 0_b|0'_b\rangle={\rm Det}[\bar{U}]^{-1/2}$  is a
normalization factor and $Z$ a symmetric matrix. The associated RPA spin state
can then be defined as
\begin{eqnarray}|0_{\rm RPA}\rangle&=&C_s\exp[\half\sum_{i\neq j}
\frac{{Z}^{ij}}{2\sqrt{s_is_j}}s_{i+}s_{j+}]|0\rangle\label{sts}\,.
\end{eqnarray}
The expectation values generated by (\ref{sts}) will be close to those obtained
with the mapping (\ref{sb}), coinciding exactly up to second order in $V$ 
(Appendix \ref{ApA}). In contrast with $|0\rangle$, the
state (\ref{sts}) is {\it entangled} (unless $V\neq 0$).

Let us note that for the quadratic Hamiltonian (\ref{H}): \\
i) $|0_{\rm RPA}\rangle =|0\rangle$ {\it if and only if} $|0\rangle$ is an
exact eigenstate of $H$, since  ${H}^b$ contains the {\it exact} matrix
elements connecting $|0\rangle$ with the rest of the Hilbert space:
\begin{equation}
{H}|0\rangle=\langle H\rangle_0|0\rangle-\half\sum_{i,j}
\Delta_-^{ij}|1_i1_j\rangle\label{eqx}\,,
\end{equation}
where $|1_i1_j\rangle=\frac{s_{i+}s_{j+}}{2\sqrt{s_is_j}}|0\rangle$ and 
we have used the mean field condition $\langle 1_i|H|0\rangle=\langle
1_i|{h}_i|0_i\rangle=0$ (Eqs.\ (\ref{h})--(\ref{mf})). Thus, if $|0_{\rm
RPA}\rangle=|0\rangle$, $Z=0$ and hence $V=0$ in ${\cal W}$, implying
$\Delta_-=0$. $|0\rangle$ is then an exact eigenstate by Eq.\ (\ref{eqx}).
Conversely, if $|0\rangle$ is an exact eigenstate, it is a solution of the mean
field equations leading to $\Delta_-=0$, implying $|0_{\rm
RPA}\rangle=|0\rangle$ (although $\Delta_+$ may be non-zero and
$\omega^\alpha\neq\lambda^\alpha$). In particular, when ${H}$ has an exactly
separable ground state $|0\rangle$ (i.e., at the factorizing field 
\cite{KTM.82,GAI.08,RCM.08}), $|0_{\rm RPA}\rangle=|0\rangle$.

ii) $|0_{\rm RPA}\rangle$ is always exact for sufficiently strong fields
($|\bm{B}|\gg J$). In this limit $|0\rangle$ is the state with all spins
$\bm{s}_i$ fully aligned along $-\bm{B}^i$ plus small corrections
($\bm{\lambda}^{i}\approx \bm{B}^i+sJ\cdot\bm{B}^i/|\bm{B}|^i$). Up to first
order in $\Delta_{\pm}/\lambda$, Eqs.\ (\ref{MH})--(\ref{stb}) lead then to
$Z^{ij}\approx V_{ij}\approx\frac{\Delta_-^{ij}}{\lambda_{i}+\lambda_{j}}$,
entailing
\begin{equation}
|0_{\rm RPA}\rangle\approx |0\rangle+\sum_{i<j}\frac{\Delta_{-}^{ij}}
{\lambda_{i}+\lambda_{j}}|1_i1_j\rangle\,, \label{mos}
\end{equation}
which, by Eq.\ (\ref{eqx}), is just the first order expansion (in
$\Delta_-/\Lambda$) of the exact ground state. 

In the case of a symmetry-breaking mean field, the RPA spin state allows to
implement the necessary rotations for symmetry restoration: The exact ground
sate will actually be close to the superposition with the correct symmetry of
the degenerate RPA ground states (rather than to a particular RPA state), as
will be discussed in Sec.\ \ref{XYZc} in the context of parity-breaking. This
restoration enlarges considerably the capabilities of the RPA.  

\subsection{Bosonic evaluation of subsystem entropy and negativity \label{bos}}

The direct evaluation of many-body correlations and entanglement measures from
the RPA spin state (\ref{sts}) is in general difficult. However, the values of 
these quantities in the associated bosonic vacuum (\ref{stb}), which will be 
close to those obtained from (\ref{sts}), can be straightforwardly evaluated 
using the general gaussian state formalism \cite{AEPW.02,ASI.04}. The reduced 
density matrix of any subsystem is just a gaussian state, i.e., a canonical 
thermal state of an effective quadratic bosonic Hamiltonian, since Wick's theorem 
holds for the evaluation of the mean value of any observable, and in particular 
those of the subsystem. We may then evaluate its entropy and other invariants 
through standard expressions for independent boson systems.

Let us formalize the previous scheme. We will use a generalized contraction 
matrix formalism, equivalent to that based on covariance matrices 
\cite{AEPW.02,ASI.04}, which is more natural for the present RPA
formulation. In the new vacuum $|0'_b\rangle$,
$\langle{{b}'}^\dagger_\alpha{b}'_{\alpha'} \rangle_{0'}=\langle {b}'_\alpha
{b}'_{\alpha'}\rangle_{0'}=0$, implying
\begin{subequations}\label{vmbs}
\begin{eqnarray}
F_{ij}&\equiv&
\langle {b}^\dagger_{j} {b}_{i}\rangle_{0'}=(VV^\dagger)_{ij},\label{Fp}\\
G_{ij}&\equiv&
\langle {b}_{j}{b}_{i}\rangle_{0'}=(VU^{\rm tr})_{ij}\,.\label{Fm}
\end{eqnarray}
\end{subequations}
Eqs.\ (\ref{sb})--(\ref{vmbs}) determine the basic RPA spin averages and
correlations, i.e.,  $\langle s_{i\mu}\rangle_{0'}=\delta_{\mu
z}(F_{ii}-s_i)$ and, for $i\neq j$,
\begin{equation}
\langle s_{i+}s_{j-}\rangle_{0'}=2\sqrt{s_is_j}\,F_{ji}\,,\;
\langle s_{i-}s_{j-}\rangle_{0'}=2\sqrt{s_is_j}\,G_{ji}\,,
\label{srpa}
\end{equation}
with $\langle s_{i\pm}s_{jz}\rangle_{0'}=0$, which coincide exactly with the
averages derived from (\ref{sts}) up to second order in $V$, i.e., first order
in the average occupation $VV^\dagger$ (normally very small outside critical
regions). Through the use of Wick's theorem, we also obtain $\langle
s_{iz}s_{jz}\rangle_{0'}= \langle s_{iz}\rangle_{0'}\langle
s_{jz}\rangle_{0'}+|F_{ij}|^2+|G_{ij}|^2$ for $i\neq j$.

We may now define the generalized contraction matrix
\begin{eqnarray}
{\cal D}&\equiv&\langle {\cal Z} {\cal Z}^\dagger\rangle_{0'}-{\cal M}=
\left(\begin{array}{cc}F&G\\\bar{G}&I+\bar{F}\end{array}\right)\,,
 \label{D}\end{eqnarray}
which exhibits the correct transformation rule under Bogoliubov
transformations: If ${\cal Z}={\cal W}{\cal Z}'$, then
\begin{eqnarray}
{\cal D}&=&{\cal W}{\cal D}'{\cal W}^\dagger \label{vmb2}
 \,.\end{eqnarray}
with ${\cal D}'=\langle {\cal Z}'{{\cal Z}'}^{\dagger}\rangle_{0'}-{\cal M}$.
Eq.\ (\ref{vmbs}) can in fact be written in the form (\ref{vmb2}) if ${\cal W}$
is the diagonalizing Bogoliubov matrix (\ref{W}) and ${\cal D'}$ the vacuum
density $(F'=G'=0)$. We may then also obtain ${\cal W}$ and ${\cal D}'$ through
the symplectic diagonalization of ${\cal D}$, i.e., through the diagonalization
of
\begin{equation}
{\cal D}{\cal M}=\left(\begin{array}{cc} F&-G\\
\bar{G}&-I-\bar{F}\end{array}\right)
\end{equation}
such that ${\cal W}^{-1}{\cal D}{\cal M}{\cal W}={\cal D}'{\cal M}$, with
${\cal D}'$ diagonal.

Let us consider now a subsystem $A$ of $m<n$ sites. It will be characterized by
a truncated contraction matrix
\begin{eqnarray}
{\cal D}_A&=&\langle {\cal Z}_A{\cal Z}^\dagger_A\rangle_{0'_b}-{\cal M}_A
=\left(\begin{array}{cc}F_A&{G}_A\\
\bar{G}_A&I+\bar{F}_A\end{array}\right)\label{DA}
 \end{eqnarray}
where ${\cal Z}_A$ contains just the bosons of sites in $A$. A symplectic
diagonalization of ${\cal D}_A$ will lead to
\begin{eqnarray} {\cal
D}_A&=&{\cal W}_A{\cal D'}_A{\cal W}^\dagger_A,\;\; {\cal
D}'_A=\left(\begin{array}{cc}{f_A}&0\\0&I+{f_A}\end{array}\right)\,,
\end{eqnarray}
where  $f_A^{\alpha\alpha'}=f_A^\alpha\delta^{\alpha\alpha'}$ with
$f_A^\alpha=\langle {b'}^\dagger_{\alpha_A}b'_{\alpha_A}\rangle_{0'}\geq 0$
(${\cal D}_A{\cal M}_A$ has eigenvalues $f_A^\alpha$ and $-1-f_A^{\alpha}$) and
${\cal W}_A{\cal M}_A{\cal W}_A^\dagger={\cal M}_A$, with ${\cal Z}_A={\cal
W}_A {\cal Z}'_A$.  The entanglement between $A$ and its complement $\bar{A}$
is then given by the associated bosonic entropy,
\begin{eqnarray}
S(\rho_A^b)&=&-{\rm Tr}\rho_A^b\log_2\rho_A^b\\
&=&-\sum_\alpha f^\alpha_A\log_2 f^\alpha_A-(1+f^\alpha_A)
\log_2(1+f^\alpha_A)\,.\label{Sblk}
\end{eqnarray}
Here $\rho_A^b\equiv {\rm Tr}_{\bar{A}}|0'_b\rangle\langle 0'_b|$ is the
bosonic reduced density of subsystem $A$, which can be explicitly written as
\begin{eqnarray}
\rho_A^b &=&C\exp[-\half {\cal Z}^\dagger_A {\cal H}_A{\cal Z}_A]
=C\exp[-\sum_\alpha \omega^\alpha_A {{b}'}^\dagger_{\alpha_A}
{b}'_{\alpha_A}]\label{rhoab}
 \end{eqnarray}
where $C={\prod_\alpha(1+f^A_\alpha)}$ and ${\cal H}_A$, ${\cal D}_A$ are
related by
\begin{equation}
{\cal D}_A{\cal M}_A=[\exp({\cal M}_A{\cal H}_A)-I]^{-1}\,.
\end{equation}
Here ${\cal H}_A$ represents an
effective ``Hamiltonian'' matrix for subsystem $A$ with symplectic eigenvalues
$\omega^\alpha_A$ such that $f^\alpha_A=(e^{\omega^\alpha_{A}}-1)^{-1}$ (and
hence $-1-f^\alpha_A=(e^{-\omega^\alpha_A}-1)^{-1}$). Eq.\ (\ref{rhoab}) leads
then to the contraction matrix (\ref{DA}), and hence to the same expectation 
values as the full vacuum $|0'_b\rangle\langle 0'_b|$ for any operator of 
subsystem $A$.

Eq.\ (\ref{Sblk}) provides a tractable RPA estimation of the entanglement
entropy of any subsystem. It is shown in the Appendix \ref{ApA} that a direct
spin evaluation of the subsystem entropy based on the RPA state (\ref{sts})
coincides with (\ref{Sblk}) up to second order in $V$.

On the other hand, the internal entanglement of subsystem $A$ with respect to a
partition $(B,C)$ of $A$ (where the complement $\bar{A}$ plays the role of an
environment) can be measured through the corresponding negativity
\cite{VW.02}, defined as minus the sum of the negative eigenvalues of
the partial transpose $\rho_{A}^{t_C}$ of $\rho_A$:
\begin{equation}
N_{BC}=\half({\rm Tr}|\rho_{A}^{t_C}|-1)\,.\label{NBC}
 \end{equation}
Expectation values with respect to $(\rho_A^b)^{t_C}$ of an observable $O_A^b$
correspond to those of the partial transpose $(O_A^b)^{t_C}$ with respect to
$\rho_A^b$. This implies the replacements $F_{ij}\leftrightarrow G_{ij}$,
$F_{j'j}\leftrightarrow {F}_{jj'}$, $G_{j'j}\leftrightarrow \bar{G}_{j'j}$, in
the contraction matrix for $j,j'\in C$, $i\in B$, leading to a matrix
$\tilde{\cal D}_A$ with symplectic eigenvalues $\tilde{f}_A^\alpha$. The latter
can now be negative. We may then still write $(\rho_A^b)^{t_C}$ as in Eq.\
(\ref{rhoab}) in terms of an effective matrix $\tilde{\cal H}_A$ with
symplectic eigenvalues $\tilde{\omega}_A^\alpha$ such that
$\tilde{f}_A^\alpha=(e^{\tilde{\omega}_A^\alpha}-1)^{-1}$.

Since the trace remains unchanged (${\rm Tr}\,(\rho_A^b)^{t_C}=1$),
$|e^{-\tilde{\omega}_A^\alpha}|<1$, implying  $\tilde{f}_A^\alpha>-1/2$.  A
negative $\tilde{f}_A^\alpha>-1/2$ corresponds to $e^{-\tilde{\omega}_A^\alpha}
<0$ and hence to a non-positive $(\rho_A^b)^{t_C}$, indicating an entangled
$\rho_A^b$ with respect to this bipartition. We then obtain, noting that
$(1+e^{-\tilde{\omega}_A^\alpha})^{-1}=(1+\tilde{f}_A^\alpha)/(1+
2\tilde{f}_A^\alpha)$, the final result \cite{VW.02,AEPW.02,ASI.04}
\begin{equation}
{\rm Tr}|(\rho_A^b)^{t_c}|=\prod_{\tilde{f}_A^\alpha<0}
\frac{1}{1+2\tilde{f}_A^\alpha}\,,\label{neg}
\end{equation}
which allows the evaluation of the negativity (\ref{NBC}). Negativities
obtained from the spin density matrices coincide with this result up to first
order in $V$ (Appendix \ref{ApA}).

In the case of a global bipartition $(A,\bar{A})$, $N_{A\bar{A}}$ becomes a
function of the reduced density $\rho_A$, namely \cite{NC.06}
\begin{equation}
N_{A\bar{A}}=\half({\rm Tr}||0\rangle\langle 0|^{\rm t_{\bar{A}}}|-1)=
\half[({\rm Tr}\sqrt{\rho_A})^2-1]\,.\label{Ng0}
\end{equation}
In a boson system, this implies that $N_{A\bar{A}}$, a limit case of Eqs.\
(\ref{NBC})--(\ref{neg}), can be also expressed just in terms of the symplectic
eigenvalues $f_A^\alpha$ of the contraction matrix ${{\cal D}_A}$:
\begin{equation}
N_{A\bar{A}}=
\half[\prod_\alpha(\sqrt{f_A^\alpha}+\sqrt{1+f_A^\alpha})^2-1]\,.\label{Ng}
\end{equation}

\subsection{Translationally invariant systems\label{TI}}
The only quantities required in the bosonic RPA scheme are, therefore, the
basic contractions (\ref{vmbs}). Their evaluation becomes remarkably simple in
translationally invariant systems, either in one or $d$ dimensions, i.e.,
systems with a common spin $s_i=s$ in a uniform field ${\bm B}^i={\bm B}$ with
couplings dependent just on separation:
\begin{equation}{J}^{i\mu\,j\nu}=J^{\mu\nu}(i-j)
 \label{JTI}
\end{equation}
where ${J}^{\mu\nu}(l)={J}^{\nu\mu}(-l)$, and
${J}^{\mu\nu}(-l)={J}^{\mu\nu}(n-l)$ in a finite cyclic chain or system (in $d$
dimensions, $i,j,l,n$ stand for $d$-dimensional vectors). We will also assume a
uniform mean field $\bm{\lambda}^i=\bm{\lambda}$, which should then satisfy
\begin{equation} \lambda^\mu=B^\mu-\sum_\nu J^{\mu\nu}_0\langle s_\nu\rangle_0
\,,\;\;J^{\mu\nu}_0\equiv\sum_l {J}^{\mu\nu}(l)\,,\label{mfti}
\end{equation}
with $\langle\bm{s}\rangle_0=-s\bm{\lambda}/\lambda$ (Eq.\ (\ref{mf})). The
uniform mean field is thus determined just by the total strengths
$J_0^{\mu\nu}$.

Choosing again the $z$ axis in the direction of $\bm{\lambda}$, such that
$\langle s_{i\mu}\rangle=-s\delta_{\mu z}$ and $B^\mu+sJ^{\mu\,z}_0=\lambda
\delta^{\mu z}$, with $\lambda>0$, the bosonized Hamiltonian will have the form
(\ref{Hb}) with couplings $\Delta_{\pm}^{ij}=\Delta_{\pm}(i-j)$. By means of a
discrete Fourier transform of the boson operators, we can rewrite it as
\begin{eqnarray}
H_b&=&\langle H\rangle_0+\sum_{k}(\lambda-\Delta_+^k)b^\dagger_k b_k
-\half(\Delta_-^k b^\dagger_k b^\dagger_{-k}+h.c.)\label{Hk}\\
{\Delta}_{\pm}^k&=&\sum_{l=0}^{n-1} e^{i2\pi kl/n}\Delta_{\pm}(l)\,,\label{Dk}
\end{eqnarray}
where $k=0,\ldots,n-1$ and $b_k=\frac{1}{\sqrt{n}}\sum_{j=1}^n e^{i2\pi
kj/n}{b}_{j}$ are boson operators in momentum space, with $b_{-k}=b_{n-k}$.
Diagonalization of (\ref{Hk}) is straightforward and leads to
\begin{equation}
H^b=\langle H\rangle_0+
\sum_k\,\omega^k {b'}^\dagger_kb'_k+\half(\omega^k-\lambda+\Delta_+^k)\,,
\end{equation}
where $\omega^k=\tilde{\omega}^k-\half(\Delta_+^k-\Delta_+^{-k})$,
${b'}^\dagger_k=u_k{b}^\dagger_k+\bar{v}_{k}b_{-k}$ and
\begin{eqnarray}
\tilde{\omega}^k&=&\sqrt{(\lambda-\tilde{\Delta}_+^k)^2-|\Delta_-^k|^2}
\,,\label{twk}\\
u_k&=&\sqrt{\frac{\lambda-\tilde{\Delta}_+^{k}
+\tilde{\omega}_k}{2\tilde{\omega}^k}}\,,\;
v_k=\frac{\Delta_-^k}{|\Delta_-^k|}
\sqrt{\frac{\lambda-\tilde{\Delta}_+^k-\tilde{\omega}_k}
{2\tilde{\omega}^k}}\,,
\end{eqnarray}
with $\tilde{\Delta}_+^k=\half(\Delta_+^k+\Delta_+^{-k})$, $u_k^2-|v_k|^2=1$,
and $u_k=u_{-k}$, $v_k=v_{-k}$.  All $\omega^k$ should be real and positive for
a stable mean field, implying the stability conditions
\begin{equation}
0\leq |\Delta_-^k|<\lambda-\Delta_+^k\,,\;\;k=0,\ldots,n-1\,.
\end{equation}
We can now obtain the basic contractions explicitly,
\begin{eqnarray}
\langle b^\dagger_{k}b_{k'}\rangle_{0'}&=&\delta_{kk'}|v_k^2|\,,\;
 \langle b_kb_{-k'}\rangle_{0'}=\delta_{kk'}u_kv_{k}= \frac{\Delta_-^{k}}
{2\tilde{\omega}^k}\,,\label{bkbk}
\end{eqnarray}
which lead finally to (Eq.\ (\ref{vmbs}))
\begin{subequations}\label{Fijl}
\begin{eqnarray}
F_{ij}=F(i-j)=\frac{1}{n}\sum_{k} e^{-i2\pi k(i-j)/n}
 |v_k^2|\,,\label{Fij}\\
 G_{ij}=G(i-j)=\frac{1}{n}\sum_{k} e^{-i2\pi k(i-j)/n}
 u_kv_k\,.\label{Gij}
\end{eqnarray}
\end{subequations}
For strong fields $|B|$ such that $\lambda\gg
|\Delta_{\pm}|$, $u_kv_k\approx \half \Delta_-^k/\lambda$ and
$|v_k^2|\approx\frac{1}{4}|\Delta_-^k|^2/\lambda^2$.
The RPA vacuum (\ref{stb}) becomes
\begin{equation}
|0'_b\rangle=C_b\exp[\half\sum_{i,j}Z(i-j){b}^\dagger_{i}{b}^\dagger_{j}]
|0_b\rangle\,,\label{st3}
\end{equation}
where $C_b=\prod_k u_k^{-1/2}$ and $Z(l)=\frac{1}{n}\sum_{k}e^{-i2\pi l k/n}
\frac{v_k}{u_k}$.

Thus, these systems allow an analytic evaluation of the contractions
(\ref{vmbs}). Both the mean field equations (\ref{mfti}) and the RPA
Hamiltonian (\ref{Hk}) become independent of the common spin $s$ after a
rescaling $J^{\mu\nu}(l)\rightarrow J^{\mu\nu}(l)/s$, which we will adopt in
what follows and which indicates that RPA is describing the large spin limit of
the system, as is apparent from Eq.\ (\ref{sb}).

\subsection{XYZ systems \label{XYZc}}
Let us now examine in more detail the previous formalism in a
translationally invariant spin $s$ array with $XYZ$ couplings of
arbitrary range in a uniform transverse
field:
\begin{equation}
H=B\sum_i s_{iz}-{\textstyle\frac{1}{2s}}\sum_{i\neq j}
\sum_{\mu=x,y,z}J_\mu(i-j)s_{i\mu}s_{j\mu}\,. \label{XYZ}
\end{equation}
Eq.\ (\ref{XYZ}) commutes with the $S_z$ spin parity,
\[ [H,P_z]=0,\;\;P_z=\exp[i\pi \sum_i(s_{iz}-s)]\,,\]
for any value of its parameters, such that the exact ground state in a finite
array will always have a definite parity outside degeneracy points. We will
focus here on the ferromagnetic type case where $J_x(l)\geq 0$ $\forall$ $l$
with
\begin{equation}|J_y(l)|\leq J_x(l)\label{fm}\,,\end{equation}
which exhibits a normal and parity breaking phase at the mean field level.

\subsubsection{RPA around the normal state}
For the Hamiltonian (\ref{XYZ}), the state $|0\rangle$ with all spins fully
aligned along the $-z$ axis is always a solution of the mean field equation
(\ref{mfti}), being the lowest solution for a sufficiently strong field $B$. It
leads to $\lambda^i=\lambda\delta^{\mu z}$, with
\begin{equation}
 \lambda=|B|+J^0_z>0\,,\;\;J^0_z\equiv\sum_l J_z(l)\,.
\end{equation}
All previous equations can then be directly applied. Now
$\Delta_{\pm}(l)=\frac{J_x(l)\pm J_y(l)}{2}=\Delta_\pm(-l)$, implying
$\Delta_{\pm}^{k}=\Delta_{\pm}^{-k}$ and 
 \begin{eqnarray}
{\omega}^k&=& \sqrt{(\lambda-J^k_x)(\lambda-J^k_y)}\,, \label{wks}
\end{eqnarray}
where $J_\mu^k=\sum_{l}e^{i2\pi kl/n}J_ \mu(l)$ ($\Delta_{\pm}^k=
\frac{J_x^k\pm J_y^k}{2}$). This solution is therefore stable provided
$J_\mu^k\leq \lambda$ $\forall k$ and $\mu=x,y$, i.e. for $|B|$ above a certain
critical field $B_c$. In the case (\ref{fm}), the strongest condition is
obtained for $k=0$, i.e.,
\begin{equation}
|B|>B_c\equiv J_x^0-J_z^0\label{Bc}\,.
\end{equation}
\subsubsection{RPA around the parity breaking state}
For $|B|\leq B_c$, the normal state becomes unstable: the lowest normal RPA
frequency $\omega^0$ vanishes for $|B|\rightarrow B_c$ and becomes imaginary
for $|B|<B_c$. The lowest mean field for $|B|<B_c$ corresponds instead to a
parity-breaking state with all spins aligned along an axis in the $xz$ plane
forming an angle $\theta$ with the $z$ axis:
\begin{equation}
|0\rangle\rightarrow |\Theta\rangle\equiv|\theta_1\ldots \theta_n\rangle\,,\;\;
|\theta_j\rangle=\exp[-i\theta s_{jy}]|0_j\rangle\,.
\end{equation}
This leads to $\langle
\bm{s}_j\rangle_0=-s(\sin\theta,0,\cos\theta)=-s\bm{\lambda}/\lambda$, with
\begin{equation}
\lambda=J_x^0,\;\;\cos\theta=B/B_c\,,\label{cost}
\end{equation}
as determined by (\ref{mfti}). We should now express the original spin
operators in terms of the rotated operators, i.e.,
\begin{equation}
s_{ix}=s_{ix'}\cos\theta+s_{iz'}\sin\theta,
\;\;s_{iz}=s_{iz'}\cos\theta-s_{ix'}\sin\theta \label{rot}
 \end{equation}
with $s_{iy}=s_{iy'}$. The RPA around this state amounts therefore to the
replacements
\begin{eqnarray}
\lambda&\rightarrow& J_x^0,\;\;J_x^k\rightarrow {J'}_x^k=
J_x^k\cos^2\theta+J_z^k\sin^2\theta\,,\label{Jxk}
\end{eqnarray}
in Eq.\ (\ref{wks}), with $J_y^k$ unchanged and
$\Delta_{\pm}^k=\half({J'}_x^k\pm J_y^k)$.

Correlations $\langle{s}_{i\mu'}s_{j\mu'}\rangle_{\rm RPA}$ of rotated spin
operators have the same previous expressions (\ref{vmbs}), whereas those of the
original operators must be obtained using Eqs.\ (\ref{rot}). It should be
remarked, however, that in a finite system, the associated RPA spin state will
no longer be a good approximation to the actual ground state due to parity
breaking. Parity restoration, at least approximately, must be implemented
before obtaining final results. We will not discuss here the case of a
continuous broken symmetry (arising for instance in the $XXZ$ case), which can
be treated through the RPA formalism of ref.\ \cite{CMR.07}.

\subsubsection{Definite Parity RPA ground states\label{PB}}
Since $[H,P_z]=0$, the parity breaking mean field state $|\Theta\rangle$ is
degenerate: Both $|\Theta\rangle$ and
$|-\Theta\rangle=P_z|\Theta\rangle$ are mean field ground states. In order to
describe the definite parity ground states, the correct RPA ground state should
be taken as the definite parity combinations 
\begin{equation}|\Theta^{\pm}_{\rm RPA}\rangle
=\frac{|\Theta_{\rm RPA}\rangle\pm|-\Theta_{\rm RPA}\rangle} {\sqrt{2(1\pm
\langle -\Theta_{\rm RPA}|\Theta_{\rm RPA}\rangle)}}\,,\label{stp}
 \end{equation}
where $|\pm\Theta_{\rm RPA}\rangle$ are the RPA states around each mean field.
The overlap $\langle-\Theta_{\rm RPA}|\Theta_{\rm RPA}\rangle= \langle
\Theta_{\rm RPA}|P_z|\Theta_{\rm RPA}\rangle$ is proportional to the overlap
between the two mean fields,
\begin{equation}
\langle-\Theta|\Theta\rangle=\cos^{2ns}\theta=(B/B_c)^{2ns}\label{ov}\,,
\end{equation}
which is small except for $B\rightarrow B_c$ or small $ns$.

Neglecting the previous overlap, Eq.\ (\ref{stp}) will lead to reduced
densities
\begin{equation}
\rho_A^\pm\approx\half [\rho_A(\theta)+\rho_A(-\theta)]\label{rhop}
\end{equation}
provided the complementary overlap $\langle -\Theta_{\rm RPA}^{\bar{A}}
|\Theta_{\rm RPA}^{\bar{A}}\rangle\propto(\frac{B}{B_c})^{2(n-n_A)s}$ can also
be neglected. Here $\rho_A(\pm\Theta)$ are the reduced spin densities
determined by each RPA state, given up to $O(V^2)$ by the expressions of
Appendix \ref{ApA}.

The restoration (\ref{rhop}) is essential to achieve a good description of the
actual subsystem entropy, although its main effect for a not too small
subsystem $A$ is actually quite simple: If the product $\rho_A(\Theta)
\rho_A(-\Theta)\propto(B/B_c)^{2n_As}$ can be neglected, Eq.\ (\ref{rhop}) can
be considered as the sum of two densities with orthogonal support and identical
distributions, leading to
\begin{equation}
S(\rho_A^\pm)\approx S(\rho_A(\theta))+1\,,\label{Sp}
\end{equation}
where $S(\rho_A(\Theta))$ can be evaluated through the boson approximation
(\ref{Sblk}). Under the same assumptions, the effect on the global negativity
(\ref{Ng0}) is just
\begin{equation}
N_{A\bar{A}}(\rho_A^{\pm})\approx 2N_{A\bar{A}}(\rho_A(\theta))+\half\,,
\label{Np}
\end{equation}
as ${\rm Tr}\sqrt{\rho_A^\pm}\approx\sqrt{2}{\rm Tr}\sqrt{\rho_A(\theta)}$,
while the subsystem negativity $N_{BC}$ of a bipartition $(B,C)$ of $A$ remains
approximately unchanged: $N_{BC}(\rho_A^\pm)\approx N_{BC}(\rho_A(\theta))$.

When the product $\rho_A(\Theta)\rho_A(-\Theta)$ cannot be neglected (as in a
subsystem of two spins), we should in principle construct the spin density
(\ref{rhop}). This can be done by rotating $\rho_A(\theta)$ (Eq.\ (\ref{rhoa})
in the mean field frame) to the original $z$ axis and removing all parity
breaking elements (which is the final effect of Eq.\ (\ref{rhop})). For
instance, the reduced two-spin density for $s=1/2$ has the blocked form
(\ref{rhoa}) in the standard basis of $s_{iz}s_{jz}$ eigenstates in the normal
phase as well as in the parity breaking phase after parity restoration
\cite{RCM.08}. The final effect on $S(\rho_A)$ is the replacement of
the term $+1$ in (\ref{Sp}) by the entropy of the reduced mean field mixture
$-\sum_{\nu=\pm} q_\nu \log_2 q_\nu$, with $q_\pm=\half(1\pm
(B/B_c)^{2s_A})$, plus small RPA corrections.

While $\rho_A^{\pm}$ are both identical in the approximation (\ref{rhop}), the
actual $\rho_A^{\pm}$ in a small system will depend on parity. The correct
parity in such a case should be chosen as that leading to the lowest energy
$E_{\rm RPA}^\pm=\langle\Theta_{\rm RPA}^\pm|H|\Theta_{\rm RPA}^\pm\rangle$.

\subsubsection{Factorizing Field}
The explicit value of the basic RPA couplings $\Delta_\pm^k$ in the parity
breaking phase are, using Eqs.\ (\ref{Jxk})--(\ref{cost}),
\begin{equation}
\Delta_\pm^k= \half[(J^k_x-J^k_z)(B/B_c)^2+J^k_z\pm J^k_y)]
\end{equation}
In the case of a common anisotropy, such that the ratio
\begin{equation}
\chi=\frac{J_y(l)-J_z(l)}{J_x(l)-J_z(l)}
\end{equation}
is {\it independent} of the separation $l$, we have
$J_y^k-J_z^k=\chi(J_x^k-J_z^k)$ and hence
$\Delta_-^k=\half(J^k_x-J^k_z)[(B/B_c)^2-\chi]$. It is then seen that if
$\chi\in[0,1]$, $\Delta_-^k=0$ $\forall$ $k$ when
\begin{equation}|B|=B_s\equiv B_c\sqrt{\chi}\label{bse}\end{equation}
with all $\Delta_-^k$ changing sign at $|B|=B_s$. Here $B_s$ is the {\it
factorizing field} \cite{AF.08,KTM.82,GAI.08,RCM.08,RCM.09}: At $B=B_s$ 
the parity breaking 
mean field state becomes an {\it exact} ground state, since the RPA corrections
vanish (sec.\ \ref{IIb}). This effect is independent of the number of spins $n$
(as long as $\chi$ is constant) and spin $s$ (with the present scaling).
Nonetheless, the actual side limits at $B=B_s$ will be given by the definite
parity states (\ref{stp}), which are still entangled. As a consequence, the
subsystem entropy $S(\rho_A)$ and the negativity $N_{A\bar{A}}$ will actually
approach a {\it finite} value for $B\rightarrow B_s$ ($1$ and $1/2$
respectively in the approximation (\ref{Sp})--(\ref{Np})), while the
entanglement between two spins will reach there infinite
range \cite{AA.06,RCM.08,MRC.08}. Note finally that at $B=B_s$, 
$\Delta_+^k=J^k_y$ and hence,
\begin{equation}
\omega^k=J_x^0-J^k_y\,.\label{omg}
\end{equation}

\section{Application\label{III}}
\subsection{Spin $s$ pair}
As a first example, let us consider a system of two spins $s$ coupled through
the Hamiltonian (\ref{XYZ}). We can obviously always set here $J_x\geq |J_y|$
(Eq.\ (\ref{fm})), since the sign of $J_x$ can be changed by a $\pi$-rotation
around the $z$ axis of one of the spins (and we can always set $|J_x|\geq|J_y|$
by a proper choice of axes). The Fourier transform of
$J_\mu(l)=\delta_{l1}J_\mu$ reduces here  to $J_\mu^k=(-1)^k J_\mu$, $k=0,1$,
leading to an attractive and a repulsive normal mode:
\[\omega_0=\sqrt{(\lambda-J_x)(\lambda-J_y)}\,,
 \;\omega_1=\sqrt{(\lambda+J_x)(\lambda+J_y)}\,.\]
The contractions (\ref{Fijl})  become $F_{ij}=\frac{\lambda-\Delta_+}
{4\omega_0}-\frac{\lambda+\Delta_+}{4\omega_1}(1-2\delta_{ij})-
\half\delta_{ij}$, $G_{ij}=\frac{\Delta_-}{4\omega_0}+
\frac{\Delta_-}{4\omega_1} (1-2\delta_{ij})$, where $\Delta_{\pm}=\half(J_x\pm
J_y)$ and replacements (\ref{Jxk}) are to be applied for $|B|<B_c$. The ensuing
entanglement entropy of the pair in the bosonic approximation (\ref{Sblk}) is
just
\begin{eqnarray}
S(\rho_1)&=&-f\log_2 f+(1+f)\log_2(1+f)+\delta\,,\label{Sf}\\
f&=&\half(\sqrt{1+\frac{\lambda^2-\overline{\omega}^2}{\omega_0\omega_1}}-1)
\,,\;\;\overline{\omega}=\frac{\omega_0+\omega_1}{2} \label{f}
\end{eqnarray}
where $f=\sqrt{(F_{11}+\half)^2-(G_{11})^2}-\half$ is the positive symplectic
eigenvalue of the $2\times 2$ contraction matrix for one spin and $\delta=0$
($1$) for $|B|>B_c$ ($<B_c$) in the approximation (\ref{Sp}), valid for
$(B/B_c)^{2s}\ll 1$). For small $f$, we may just use $S(\rho_1)\approx f(\log_2
e-\log_2 f)$, with $f\approx F_{11}$, in agreement with the results of Appendix
A.

Thus, at the RPA level entanglement is determined by the average local
occupation $f$ and driven by the ratio $\frac{\lambda^2-
\overline{\omega}^2}{\omega_0\omega_1}$, which is small away from $B_c$ and
vanishes at $B=B_s$ (where $\overline{\omega}=\lambda=J^0_x$ by Eq.\
(\ref{omg}), and hence $f=0$). For $|B|\gg B_c$,
$f\approx(\frac{J_x-J_y}{4B})^2$, while in the vicinity of $B_s$, $f\propto
(B-B_s)^2$. For $B\rightarrow B_c$, $f\approx
\half\sqrt{\frac{\lambda^2-\overline{\omega}^2}{\omega_0\omega_1}}\propto
|B-B_c|^{-1/4}$, with  $S(\rho_1)\approx\log_2 f e$.
\begin{figure}[t]
\centerline{\hspace*{-0.cm}\scalebox{.8}{\includegraphics{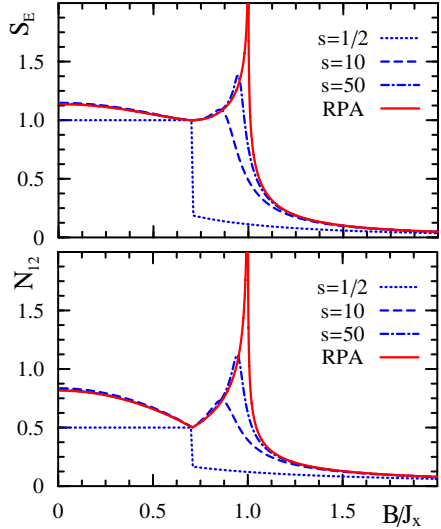}}}
\vspace*{-1cm}

\caption{Entanglement between two spins $s$ as a function of the transverse
field $B$ for an $XY$ coupling with $J_y/J_x=0.5$. The exact entanglement
entropy $S_E=S(\rho_1)$ (top) and negativity (bottom) for different values of
the spin $s$, and the bosonic RPA results, Eqs.\ (\ref{Sf}), (\ref{Nf}) are
depicted. The exact results approach those of RPA as $s$ increases, differences
for not too small $s$ arising just for $B$ close to $B_c=J_x$. At the
factorizing field $B_s\approx B_c/\sqrt{2}$, $S_E=1$ while $N_{12}=1/2$.}
\label{fig1} %\vspace*{-.5cm}
\end{figure}

The bosonic RPA negativity (\ref{NBC})-(\ref{neg}) becomes
\begin{eqnarray}
N_{12}&=&\frac{-\tilde{f}}{1+2\tilde{f}}
=f+\sqrt{f(f+1)}\label{Nf}
\end{eqnarray}
where $\tilde{f}=f-\sqrt{f(f+1)}$ is the negative symplectic eigenvalue of the
$4\times 4$ contraction matrix. Correction (\ref{Np}) ($N_{21}\rightarrow
2N_{21}+\half$) should be applied for $|B|<B_c$. For small $f$, we have simply
$N_{12}\approx -\tilde{f}\approx \sqrt{f}$. This will lead to a slope
discontinuity of $N_{12}$ at the factorizing field $B_s$ (see Fig.\
\ref{fig1}), as $f$ vanishes there quadratically ($N_{12}-\half\propto |B-B_s|$
for $B\approx B_s$).
On the other hand, for $f\rightarrow\infty$ ($|B|\rightarrow B_c$),
$\tilde{f}\rightarrow -\half$, with $\tilde{f}\approx -\half+\frac{1}{8f}$ and
$N_{12}\approx 2f$. Both $S(\rho_1)$ and $N_{12}$ are concave increasing
functions of $f$ and measure the entanglement of the pair.

Comparison with exact numerical results, obtained through the diagonalization
of $H$ (a $(2s+1)^2\times (2s+1)^2$ matrix), are shown in Fig.\ \ref{fig1} for
the $XY$ case ($J_z=0$) with anisotropy $\chi=J_y/J_x=0.5$. Exact results are
seen to rapidly approach the RPA values (\ref{Sf})--(\ref{Nf}) as the spin $s$
increases, the discrepancy for finite $s$ arising just in the vicinity of $B_c$
or for very small $s$, i.e., where tunneling effects arising from the non-zero
overlap (\ref{ov}) between the degenerate parity breaking states become
appreciable.

\begin{figure}[t]
\centerline{\hspace*{-0.2cm}\scalebox{.6}{\includegraphics{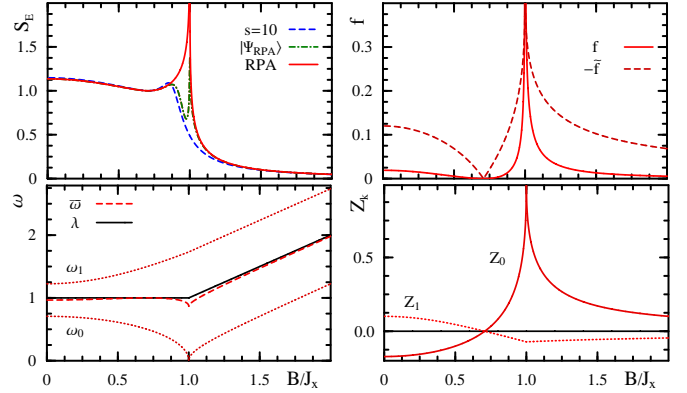}}}
\vspace*{-1cm}

\caption{Top:  Left: The entanglement entropy obtained from the definite parity
RPA spin state (\ref{stp}) (dashed-dotted line), compared with the bosonic RPA
result (\ref{Sf}) and the exact value, for $s=10$ at the same parameters of
fig.\ \ref{fig1}. The result from the RPA spin state improves the bosonic RPA
for $B$ just below $B_c$. Right: The average local boson occupation $f$, Eq.\
(\ref{f}), which is small away from $B_c$, and the negative eigenvalue
$\tilde{f}$ of the partial transpose of the contraction matrix
($\tilde{f}\approx \sqrt{f}$ for small $f$). Bottom: Left: RPA energies
$\omega_0$, $\omega_1$, together with the mean field energy $\lambda$ and the
mean RPA energy $\overline{\omega}$ appearing in (\ref{f}). Right: The
quantities $Z_k=v_k/u_k$ for $k=0,1$, which determine the RPA state (\ref{st3})
and vanish at the factorizing field $B_s$.}
 \label{fig2}%\vspace*{-.5cm}
\end{figure}

Nonetheless, this overlap can be taken into account using the full definite
parity RPA spin state (\ref{stp}) with lowest energy, which for finite $s$
improves results for $B$ close to $B_c$ (but otherwise yields results almost
coincident with those of the corrected bosonic RPA), as seen in fig.\
\ref{fig2}.  Eq.\ (\ref{stp}) also yields the {\it exact} side limits at the
factorizing field \cite{RCM.08} for {\it any} $s$, although for $\chi=0.5$
these limits rapidly approach the high spin values $S(\rho_1)=1$ and
$N_{12}=\half$ predicted by the approximations (\ref{Sp})--(\ref{Np}).

Fig.\ 2 also depicts the behavior of the average occupations $f$ and
$\tilde{f}$. The former is seen to be quite small ($f\alt 0.05$) except in the
vicinity of $B_c$, implying that away from $B_c$, all bosonic RPA results can
be reproduced by the spin densities of Appendix A, with $\tilde{f}\approx
\sqrt{f}$. In the bottom panels we depict the RPA energies $\omega_0,\omega_1$
and the RPA state coefficients $Z_k\equiv v_k/u_k$ used in Eq.\ (\ref{st3}).
Although $\omega_0$ vanishes at $B_c$, the difference
$\lambda-\overline{\omega}$, responsible for entanglement, remains everywhere
quite small. Both $Z_k$ vanish and change sign at the factorizing field $B_s$,
indicating a qualitative change in the type of correlations at this point:
Entanglement between two spins $1/2$ is well known to change from antiparallel
to parallel (in the original frame) at $B_s$ \cite{RCM.08}, an effect arising
within the RPA from this sign change.

\subsection{Fully connected spin system}
Let us now consider a fully and uniformly connected $XYZ$ array of $n$ spins,
where
\begin{equation}J_\mu(l)=(1-\delta_{l0})J_\mu/(n-1)\,,\label{jmul}
\end{equation}
in (\ref{XYZ}). This scaling ensures a finite intensive energy $\langle
H\rangle/n$ for large $n$ and finite $J_\mu$. Entanglement properties of this
well-known model \cite{RS.80,LMG.65} for $s=1/2$ in the large $n$ limit 
have been previously analyzed \cite{LORV.05}, including recently 
Holstein-Primakoff based bosonization \cite{WVB.10,VDB.07,BDV.06,DV.05}. 
Direct application of the present RPA formalism will be here shown to yield full analytic expressions for any size $n$ and spin $s$. 
The present treatment does not exactly coincide with that of refs.\ \cite{WVB.10,VDB.07}, since the absence of self-interacting terms $\propto s_{i\mu}s_{i\nu}$ (non-trivial for $s>1/2$) is here exactly taken into account and leads to repulsive RPA corrections ($\omega_1$), non-zero for finite $n$.  
The Fourier transform of (\ref{jmul}) is
$J_\mu^0=J_\mu$ and $J_\mu^k=-J_\mu/(n-1)$ for $k=1,\ldots,n-1$, leading again
to two distinct RPA energies: One associated with a fundamental attractive mode
($\omega_0$) and $n-1$ degenerate weak repulsive modes $\omega_k=\omega_1$,
$k\neq 0$, which just add a small repulsive correction accounting for the
absence of self-energy terms:
\[\omega_0=\sqrt{(\lambda-J_x)(\lambda-J_y)},
\;\omega_1=\sqrt{(\lambda+{\textstyle\frac{J_x}{n-1}})
 (\lambda+{\textstyle\frac{J_y}{n-1}})}\]
where the replacements (\ref{Jxk}) are to be used for $B<B_c$. The ensuing
contractions (\ref{Fijl})  become here obviously independent of separation for
$ i\neq j$:
\begin{subequations}\label{Fijc}
\begin{eqnarray}
F_{ij}&=&{\textstyle\frac{1}{2n}[\frac{\lambda-\Delta_+^0}{\omega_0}
-\frac{\lambda-\Delta_+^1}{\omega_1}(1-n\delta_{ij})]-\half\delta_{ij}}\,,\\
G_{ij}&=&{\textstyle\frac{1}{2n}[\frac{\Delta_-^0}{\omega_0}
-\frac{\Delta_-^1}{\omega_1}(1-n\delta_{ij})]}\,,
\end{eqnarray}\end{subequations}
and imply that for any bipartition $(L,n-L)$, the entanglement entropy
$S(\rho_L)$ will depend just on $L$. Moreover, there is again a {\it single}
non-zero eigenvalue $f_L$ of the reduced matrix ${\cal D}_L$ of $L$ spins for
{\it any} $L$ (see Appendix B), such that in the bosonic approximation
(\ref{Sblk})--(\ref{Sp}),
\begin{eqnarray}
S(\rho_L)&=&-f_L\log_2f_L+(1+f_L)\log_2(1+f_L)+\delta\,,\label{SfL}\\
\nonumber\\
f_L&=&\half[\sqrt{1+2\alpha_L\Delta}-1]\,,\;\;\;\alpha_L=L(n-L)/n^2\,,
\label{fl}
\end{eqnarray}
where $\delta=0$ ($1$) for
$|B|<B_c$ ($(B/B_c)^{2Ls}\ll 1$) and
\begin{eqnarray}
\Delta
&=&\frac{n^2(\lambda^2-\overline{\omega}^2)}{2(n-1)\omega_0\omega_1}\,,
\;\;\overline{\omega}=\frac{\omega_0+(n-1)\omega_1}{n}\,.
\end{eqnarray}
For $n=2$ we recover Eqs.\ (\ref{Sf})--(\ref{f}), while for large $n$,
$\Delta\approx \frac{\lambda-\Delta_+^0}{\omega_0}-1$. Entanglement is then
driven again by the ratio $\frac{\lambda^2-\overline{\omega}^2}
{\omega_0\omega_1}$, which is small away from $B_c$ and vanishes at $B_s$. For
small $\Delta$, $f_L\approx\half\alpha_L\Delta$, with
$\Delta\approx\half(\frac{n}{(n-1)}\frac{J_x-J_y}{2B})^2$ for $|B|\gg B_c$ and
$\Delta\propto (B-B_s)^2$ in the vicinity of $B_s$. For $B\rightarrow B_c$,
$f_L\propto \sqrt{\alpha_L}(B-B_c)^{-1/4}$ and $S(\rho_L)\approx \log_2 f_L e$.

\begin{figure}[t]
\centerline{\hspace*{-0.2cm}\scalebox{.6}{\includegraphics{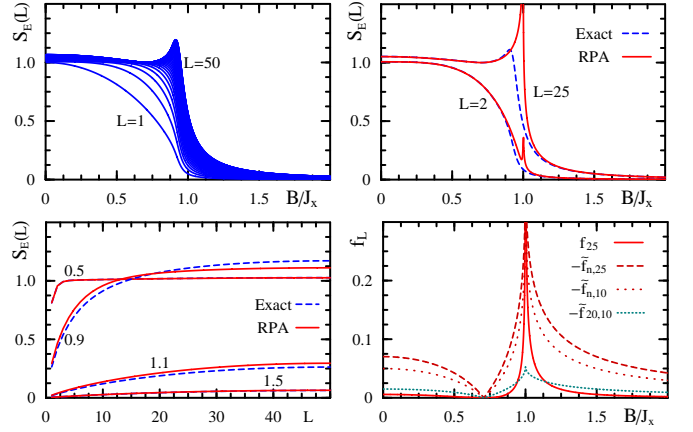}}}
\vspace*{-0.5cm}

\caption{Results for the fully connected spin $1/2$ array of $n=100$ spins.
Top: Left: Exact entanglement entropies $S_E(L)=S(\rho_L)$ of subsystems with
$L\leq n/2$ spins as a function of the magnetic field. Right: Comparison
between exact and RPA results for $S(\rho_L)$. Bottom: Left: Exact and RPA
results for $S(\rho_L)$ as a function of the subsystem size $L$ at four
different field ratios $B/B_c$. Right: Magnetic behavior of the average boson
occupation number (\ref{fl}) for $L=25$ and the negative symplectic eigenvalue
(\ref{flm}) of the partial transposed contraction matrix for different $L$,
$m$.} %\vspace*{-0.5cm}
  \label{fig3}
\end{figure}

\begin{figure}[t]
\centerline{\hspace*{-0.2cm}\scalebox{.6}{\includegraphics{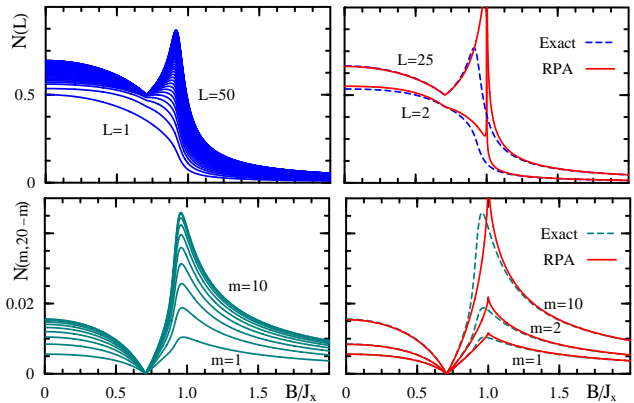}}}
\vspace*{-0.75cm}

\caption{Top: Left: Exact global negativities $N(L)=N_{L,n-L}$ between $L$ and
$n-L$ spins  in the fully connected array. Right: Comparison between exact and
RPA results for $N(L)$ for two values of $L$. Bottom: Left: Exact subsystem
negativities $N_{m,L-m}$ between $m$ and $L-m$ spins in a subsystem of $L=20$
spins. Right: Comparison between exact and RPA results for $N_{m,L-m}$.}
  \label{fig4}%\vspace*{-0.5cm}
\end{figure}

The bosonic negativity of a bipartition $(m,L-m)$ of a subsystem of $L\leq n$
spins can again be explicitly obtained, since there is also {\it a single}
negative eigenvalue $\tilde{f}_{Lm}$ of the partial transpose of the
contraction matrix (see appendix \ref{ApB}):
\begin{eqnarray}
 N_{m,L-m}&=&\frac{-\tilde{f}_{Lm}}{1+2\tilde{f}_{Lm}}\,,\label{Nlm}\\
\tilde{f}_{Lm}&=&\half\sqrt{1+\gamma_{Lm}\Delta-\sqrt{8\beta_{Lm}
\Delta+\gamma^2_{Lm}\Delta^2}}-\half\label{flm}\\
\;\;\gamma_{Lm}&=&\alpha_L+4\beta_{Lm}\,,\;\;\beta_{Lm}=m(L-m)/n^2\,.
\end{eqnarray}
For a global partition ($L=n$), $\alpha_n=0$ while $\beta_{nm}=\alpha_m$,
and
$\tilde{f}_{nL}=f_L-\sqrt{f_L(f_L+1)}$, with $N_{nL}=f_L+\sqrt{f_L(f_L+1)}$, as
in Eq.\ (\ref{Nf}). In general, for small $\Delta$,
\begin{equation}
\tilde{f}_{Lm}\approx -\sqrt{\half\beta_{Lm}\Delta}\approx
 -\sqrt{(\beta_{Lm}/\alpha_L)f_L}\label{flmp}
 \end{equation}
such that for strong fields, $\tilde{f}_{Lm}\approx
\sqrt{\beta_{Lm}}\frac{n}{n-1}\frac{J_x-J_y}{4B}$, while for $B$ close to
$B_s$, $\tilde{f}_{Lm}\propto\sqrt{\beta_{Lm}} |B-B_s|$. On the other hand, for
$B\rightarrow B_c$, $\tilde{f}_{Lm}\rightarrow -\half(1-\sqrt{\frac{\alpha_l}
{\alpha_L+4\beta_{Lm}}})+O(|B-B_c|^{1/2})$ if $\alpha_L\neq 0$, implying that
subsystem negativities $N_{m,L-m}$ with $L<n$ {\it remain finite} at $B_c$
(in agreement with the results of \cite{WVB.10}),
as $\tilde{f}_{Lm}$ remains {\it above} $-\half$.

In the parity breaking phase, the replacement (\ref{Np}) ($N\rightarrow
2N+\half$) should be used for global negativities $N_{n,L-n}$, whereas
subsystem negativities $N_{m,L-m}$ remain unchanged after parity restoration if
$(B/B_c)^{2s(n-L)}$ and $(B/B_c)^{2sL}$ can both be neglected.

Eqs.\ (\ref{SfL})--(\ref{flm}) represent essentially the exact expressions for
the subsystem entropy and negativity for large spin at finite $n$, as well as
for large $n$ at finite spin, as verified by exact numerical calculations. For
instance, exact (obtained through diagonalization of $H$) and RPA results for a
spin $1/2$ array of $n=100$ spins are shown in figs.\ \ref{fig3}--\ref{fig4}.
RPA results for the entanglement entropy are quite accurate except in the
vicinity of $B_c$, differences decreasing as $n$ or $s$ increases.  For large
$L$ they were obtained with the previous expression (\ref{SfL}) whereas for
small $L$ (like the $L=2$ case), we have used the proper spin state
(\ref{rhop}), whose main effect is to take into account the correct overlap for
$B$ below but close to $B_c$ (roughly, $\delta$ replaced by the entropy of the
reduced mean field superposition).

The variation of $S(\rho_L)$ with $L$ at fixed field (bottom left panel in
Fig.\ \ref{fig3}) is also correctly predicted, being quite accurate both in the
normal and parity breaking phase for fields not too close to $B_c$. The bottom
right panel shows that $f_L$ remains small except for $B$ around $B_c$, while
$\tilde{f}_{Lm}$ becomes also small as $L$ decreases, in full agreement with
Eq.\ (\ref{flmp}). RPA results for global ($N_{n,L-n}$)  and in particular
subsystem negativities ($N_{m,L-m}$ for $L<n$), which are much smaller and 
vanish at $B_s$, are
also very accurate, as seen in Fig.\ \ref{fig4}. Subsystem negativities were
directly obtained with Eq.\ (\ref{Nlm}), whereas global negativities were
corrected with Eq.\ (\ref{Np}) for $B<B_c$ and large $L$ and using (\ref{rhop})
for $L=2$.

\section{Discussion}
We have shown that the mean field plus RPA method is able to provide, through
the bosonic representation, a general tractable method for estimating, in the
ground state of general spin arrays, the entanglement entropy of any
bipartition of the whole system as well as the negativity associated with any
bipartition of any subsystem. The approach becomes fully analytic in systems
with translational invariance, where no numerical diagonalization is required
for obtaining the basic contraction matrices.

The bosonic treatment provides essentially the exact behavior of the system in
the large spin limit. Finite spin corrections can be taken into account through
the corresponding RPA spin state, which allows in particular to implement the
non-negligible symmetry restoration effects in the case of the parity-breaking
mean field, but which otherwise yields result which are in full agreement with
the bosonic treatment at first order in the average local boson occupation. The
latter is normally very low away from critical regions.

Through direct application of the present method, simple analytic expressions
for the entanglement entropy and negativities for a spin $s$ pair and for a
fully connected array of $n$ spins $s$ in a uniform field, have been
straightforwardly obtained, which depend explicitly on the RPA energies. 
The agreement with exact numerical results is
confirmed to improve as the spin $s$ increases at fixed size, and in the fully
connected case also as $n$ increases at fixed $s$, differences being in fact
negligible away from the critical region for not too small $s$ or size.

An important general prediction that arises from the present treatment is that
entanglement from elementary excitations approaches a non-vanishing spin
independent limit as the spin increases. An RPA quantum regime, characterized by
weak entanglement, emerges then between strictly classical and strongly quantum
regimes.
	
%\section{Acknowledgements}
The authors acknowledge support from CIC (RR) and CONICET (JMM,NC) of
Argentina.

\appendix
\section{RPA spin densities\label{ApA}}
We will here construct the spin density matrices compatible with the RPA spin
state (\ref{sts}) and the contractions (\ref{vmbs}) up to second order in $V$,
i.e., first order in the average occupation $VV^\dagger$ (implying zero or one
boson per site). At this order, $F\approx GG^\dagger$ (Eqs.\ (\ref{vmbs})) and
the support of $\rho=|0_{\rm RPA}\rangle\langle 0_{\rm RPA}|$ is just the
subspace spanned by the mean field state $|0\rangle$ plus the two site
excitations $|1_i 1_j\rangle$ (Eq.\ (\ref{eqx})), leading to
\begin{eqnarray}
  \label{eq:densitymatrixpure}
  \rho&\approx&\left(
\begin{array}{c c}
\bm{G}\bm{G}^{\dagger} &  \bm{G}\\
\bm{G}^{\dagger} &1-\bm{G}^{\dagger}\bm{G}
\end{array}
\right)
\end{eqnarray}
where $\bm{G}$ denotes a column matrix of elements $G_{ij}$, $i<j$. At this
order, $\rho^2=\rho$. The ensuing reduced density matrix $\rho_{A}={\rm
Tr}_{\bar{A}}\,\rho$ of a subsystem $A$ of $L$ spins becomes
\begin{eqnarray}
  \label{eq:subsystemdensitymatrix}
  \rho_A&\approx&\left(
\begin{array}{ccc}
\bm{G}_A\bm{G}^{\dagger}_A& 0& \bm{G}_A\\
0&F_A-G_A G_A^\dagger &0\\
\bm{G}_A^{\dagger} & 0&1-{\rm tr}\,F_A+\bm{G}_A^\dagger\bm{G}_A
 \end{array}\right) \label{rhoa}\end{eqnarray}
where $F_A$, $G_A$ are the $L\times L$ contraction matrices of subsystem $A$
and $\bm{G}_A$ the concomitant column vector (of length $L(L-1)/2$). The
central block contains  the one-site elements $|1i\rangle\langle 1_j|$ arising
from the partial trace of $\bm{G}\bm{G}^\dagger$. Here we have used the
approximate identity $\sum_{k\in\bar{A}}G_{ik}G^\dagger_{kj} \approx
F_{ij}-\sum_{k\in A}G^\dagger_{ik}G_{jk}$ for $i,j\in A$ (and neglected
diagonal elements $G_{ii}$, of higher order due to the absence of self-energy
terms), which allows to write $\rho_A$ entirely in terms of local contractions.
Eq.\ (\ref{rhoa}) is then in agreement with direct state tomography at this
order (for $i,j,k,l\in A$, $\langle b^\dagger_jb_i\prod_{k\neq
i,j}(1-b^\dagger_kb_k)\rangle_{0'} \approx (F_A-G_AG^\dagger_A)_{ij}$, $\langle
b^\dagger_i b^\dagger_jb_kb_l\rangle_{0'}\approx G_{kl}\bar{G}_{ij}$). Up to
$O(V^2)$, $\rho_A$ is a positive matrix with ${\rm Tr} \rho_A=1$, but is no
longer pure.

Its entropy $S(\rho_A)=-{\rm Tr}\rho_A\,\log_2\rho_A$ is determined, at this
order, by the central block $\rho_A^1=F_A-G_A G_A^\dagger$,
\begin{eqnarray}
S(\rho_A)&\approx &{\rm tr}\,\rho_A^1(\log_2 e-\log_2 \rho_A^1)\,,
 \label{SAa}
\end{eqnarray}
which coincides with Eq.\ (\ref{Sblk}) up to second order in $V$ (at this order
$f_A^\alpha$ coincides with the eigenvalues of $\rho_A^1$ and Eq.\ (\ref{Sblk})
becomes $\approx \sum_\alpha f_A^\alpha(\log_2 e-\log_2 f_A^\alpha$)).

On the other hand, the leading term in the negativity arising from a
bipartition $(B,C)$ of $A$ is of first order in $V$ and is just the sum of the
singular values of the submatrix $G_{BC}$ (of elements $G_{ij}$, $i\in B$,
$j\in C$), whence $N_{BC}\approx {\rm tr}\,[G_{BC}G^\dagger_{BC}]^{1/2}$. At
this order, the negative symplectic eigenvalues $\tilde{f}^\alpha_A$ in
(\ref{neg}) are again minus the singular values of $G_{BC}$, while Eq.\
(\ref{NBC}) becomes $N_{BC}\approx-\sum_\alpha\tilde{f}_A^{\bar{\alpha}}$,
leading again to the previous result.

Let us finally notice that Eq.\ (\ref{rhoa}) always commutes with the $S_z$
parity (along the mean field axis) of subsystem $A$, i.e., $[\rho_A,P_{zA}]=0$,
$P_{zA}=\exp[i\pi\sum_{i\in A}(s_{iz}-s_i)]$. In the case of two spins $i,j$,
$\bm{G}_A$ has length 1 and Eq.\ (\ref{rhoa}) is just a $4\times 4$ blocked
matrix, while in the case of a single spin $i$, $\bm{G}_A$ has length $0$ and
Eq.\ (\ref{rhoa}) becomes just $\rho_i\approx F_{ii}|1_i\rangle\langle
1_i|+(1-F_{ii})|0_i\rangle\langle 0_i|$.

\section{Fully connected system\label{ApB}}
In the fully connected $XYZ$ spin system, the contractions (\ref{Fijc}) are of
the form $F_{ij}=F_0\delta_{ij}+F_1$, $G_{ij}=G_0\delta_{ij}+G_1$, with
$F_0,F_1$, $G_0,G_1$ real. The ensuing contraction matrix ${\cal D}_L$ for a
subsystem of $L$ spins will then have symplectic eigenvalues (see also 
\cite{ASI.04}) 
\begin{eqnarray}
f_L&=&\sqrt{(F_0+LF_1+\half)^2-(G_0+LG_1)^2}-\half \label{f1a}\\
 f_0&=&\sqrt{(F_0+\half)^2-G_0^2}-\half \label{f1b}
 \end{eqnarray}
plus their partners $1+f_L$, $1+f_0$, where $f_L$ is non-degenerate while $f_0$
has $L-1$ degeneracy. Eqs.\ (\ref{f1a})--(\ref{f1b}) can be obtained either by
a Fourier transform of the local operators or by noticing that the $L\times L$
contraction matrix $F_L$ can be written as $F_L=F_0 I_L+F_1\bm{1}_L\bm{1}_L^t$
(and similarly for $G_L$), with $I_L$ the $L\times L$ identity and $\bm{1}_L$ a
column $L\times 1$ vector with unit elements. $F_L$ and $G_L$ will then be
diagonal in the same local basis with eigenvalues $F_0+LF_1$ and $F_0$ ($L-1$
degenerate) and similarly for $G_L$, which leads to Eqs.\
(\ref{f1a})--(\ref{f1b}). In the case of a global vacuum, $f_0=0$ (since for
$L=n$, we should have $f_{L=n}=f_0=0$), implying a single positive eigenvalue
$f_L$ for any $L<n$. Eq.\ (\ref{f1a}) leads then to Eq.\ (\ref{fl}).

For evaluating the negativity $N_{mp}$ of a bipartition $(m,p)$ of a subsystem
of $L$ spins ($m+p=L$), we may first note that $F_L$ will be composed of blocks
$F_{mm}=F_0I_m+ F_1\bm{1}_m\bm{1}_m^t$, $F_{mp}=F_1\bm{1}_m\bm{1}_{p}^t=
F_{pm}^t$ and $F_{pp}=F_0I_p+F_1\bm{1}_p\bm{1}_p^t$, and similarly for $G_L$. A
local transformation allows then to write $F_L$ as a direct sum
of a $(L-2)\times (L-2)$ diagonal block $F_0I_{L-2}$ plus the block $F_0
I_2+F_1(^{m\;\sqrt{mp}}_{\sqrt{mp}\;\;p})$, and similarly for $G_L$. The
ensuing partially transposed contraction matrix will then have symplectic
eigenvalues $\tilde{f}_0=f_0$ (Eq.\ (\ref{f1b})), $L-2$ degenerate (with
$\tilde{f}_0=0$ for a global vacuum) and
\begin{eqnarray}
\tilde{f}_{Lm}^\pm&=&\half\sqrt{{\rm Tr}{\cal A}^2\pm\sqrt{ ({\rm Tr}{\cal
A}^2)^2-16\,{\rm det}{\cal A}}}-\half
 \end{eqnarray}
together with their partners $1+\tilde{f}_0$, $1+\tilde{f}_{Lm}^\pm$, where
${\cal A}=(^{A_{FG}\;-A_{GF}}_{A_{GF}\;-A_{FG}})$ is a $4\times 4$ matrix with
blocks $A_{FG}=(\half +F_0)I_2+(^{mF_1\;\;\sqrt{mp}G_1}
_{\sqrt{mp}G_1\;\;pF_1})$ and similarly for ${\cal A}_{GF}$. Here
$\tilde{f}_{LM}^+>0$ but $\tilde{f}_{LM}^-<0$. The latter is the single
negative symplectic eigenvalue given in Eq.\ (\ref{flm}).

\end{document}